\documentclass[twocolumn,pra,aps,superscriptaddress,showpacs]{revtex4}
\usepackage{xspace,amsmath,amsfonts,amsthm,amssymb,amsbsy,graphicx,color}

\begin{document}

\newtheorem{theo}{Theorem}
\newtheorem{lemma}{Lemma} 

\title{Rapid-purification protocols for optical homodyning}

\author{Aravind Chiruvelli}

\affiliation{Hearne Institute for Theoretical Physics, Department of Physics and Astronomy, Louisiana State University, Baton Rouge, LA 70803, USA}

\author{Kurt Jacobs}

\affiliation{Department of Physics, University of Massachusetts at Boston, 100 Morrissey Blvd, Boston, MA 02125, USA}

\affiliation{Hearne Institute for Theoretical Physics, Department of Physics and Astronomy, Louisiana State University, Baton Rouge, LA 70803, USA}

\begin{abstract}
We present a number of rapid-purification feedback protocols for optical homodyne detection of a single optical qubit. We derive first a protocol that speeds up the rate of increase of the average purity of the system, and find that like the equivalent protocol for a non-disspative measurement, this generates a deterministic evolution for the purity in the limit of strong feedback. We also consider two analogues of the Wiseman-Ralph rapid-purification protocol in this setting, and show that like that protocol they speed up the average time taken to reach a fixed level of purity. We also examine how the performance of these algorithms changes with detection efficiency, being an important practical consideration.    
\end{abstract} 

\maketitle

\section{Introduction} 
Rapid-purification protocols increase the rate at which the state of a system is purified by a continuous measurement~\cite{rapidP, Combes06, Wiseman06x, Jordan06, Ralph06, Hill07, Griffith06x, Hill07x, Wiseman07}. They do this by applying feedback control to the system as the measurement proceeds. All such protocols described to date have been devised for  continuous measurements of an observable (that is, measurements that are not dissipative). Under this kind of measurement the evolution of the system density matrix, $\rho$, is given by the stochastic master equation~\cite{JacobsSteck06, Brun02, BelavkinLQG}, 
\begin{eqnarray}
  d\rho & = &  -(i/\hbar) [H,\rho]dt  - k[X,[X, \rho ]] dt  \nonumber \\ 
           &    &  + \sqrt{2k} ( X\rho + \rho X - 2\langle X \rangle \rho )dW ,
           \label{eq1}
\end{eqnarray}
where $X$ is the hermitian operator corresponding to the observable being measured, $H$ is the Hamiltonian of the system, $dW$ is Gaussian white noise satisfying the Ito calculus relation $dW^2 = dt$. The observers continuous measurement record, which we will denote by $r(t)$, is given by $dr = \langle X \rangle dt + dW/\sqrt{8k}$. This kind of measurement will project the system onto an eigenstate of $X$ after a time $t \gg 1/(\Delta^2k)$, where $\Delta$ is the difference between the two eigenvalues of $X$ that are nearest each other. 

Photon counting and optical homodyning do not fall into the above class of measurements because they subject the system to dissipation. Thus if one has a single optical qubit, consisting of a single mode containing no more than one photon, and one measures it with a photon counter, then regardless of whether the measurement tells us that the state was initially $|0\rangle$ or $|1\rangle$, as $t\rightarrow\infty$ the final state is always $|0\rangle$. If we wish we can think of this as a measurement of the photon number (that is, a measurement in the class above with $X = a^\dagger a$), followed by an irreversible operation that takes both $|0\rangle$ and $|1\rangle$ to the vaccum. 

Our purpose here is to examine whether there exist rapid-purification feedback protocols for homodyne detection performed on a single optical qubit, and if so, to compare their properties with those pertaining to a continuous measurement of an observable on a single qubit. Our motivation is partly theoretical interest regarding the effect of dissipation on rapid-purificaton protocols, and partly to explore whether such protocols can be implemented in an optical setting. Before we begin it is worth recalling the properties of the single-qubit rapid-purification protocols that have been derived to date for non-dissipative measurements. The first is the protocol introduced by one of us~\cite{rapidP} (see also~\cite{Wiseman07}) in which one applies feedback control to speed up the  increase in the {\em average} purity of the system. The average here is taken over all possible realizations of the measurement (all possible measurement records $r(t)$). The protocol involves applying feedback during the measurement to keep the Bloch vector of the state of the qubit perpendicular to the basis of the measured observable, $X$. In the limit of strong feedback, and high final average purity, this provides a factor of two decrease in the time required to reach a given average purity. In the limit of strong feedback the protocol also eliminates the stochasticity in the purification process, so that the purity increases deterministically. 

The second protocol, introduced by Wiseman and Ralph~\cite{Wiseman06x} (see also~\cite{Griffith06x, Wiseman07}), involves applying feedback to keep the Bloch vector parallel to the basis of the measured observable. (If the system has no appreciable Hamiltonian, then the measurement will do this of its own accord, and feedback is not required.) This protocol minimizes the {\em average time} one has to wait to reach a given purity. The decrease in this average waiting time over the previous protocol is a factor of two, and in this case the evolution of the purity is stochastic. 

In the next section we examine homodyne detection of a single optical qubit, and derive a deterministic rapid-puritifcation protocol equivalent to the first protocol discussed above. In Section~\ref{sec3} we calculate the performance of two protocols that are analogous in various ways to the Wiseman-Ralph protocol. Section~\ref{conc} summarizes with some concluding remarks. 

\section{Rapid Purification for Optical Homodyning} 
\label{sec2}
The dynamics of a single mode of an optical cavity, where the output light is monitored via homodyne detection, is given by~\cite{Wiseman93} 
\begin{eqnarray}
   d \rho & = & - \gamma D[a]\rho dt + \sqrt{2\eta \gamma} ( a e^{i\theta} \rho + \rho a^\dagger e^{-i\theta}) dW \nonumber \\
             &  &  - \sqrt{2\gamma} \langle a e^{i\theta} + a^\dagger e^{-i\theta} \rangle \rho dW ,
             \label{SMEhom} 
\end{eqnarray}
where $D[a]\rho \equiv a^\dagger a \rho + \rho a^\dagger a - 2 a \rho a^\dagger$,  $\rho$ is the state of the mode, $a$ is the mode annihilation operator, $\gamma$ is the decay rate of the mode from the cavity, and $\eta$ is the efficiency of the photodetectors. here we have moved into the interaction picture, and thus eliminated  the mode Hamiltonian $H_0 = \hbar\omega a^\dagger a$. In this case the observer's measurement record is given by $dr = \langle a + a^\dagger \rangle dt + dW/\sqrt{8\gamma}$. If the state of the mode has no more than one photon, then we can replace $a$ with the Pauli lowering operator $\sigma_- = \sigma_x - i\sigma_y$, and the SME becomes 
\begin{widetext}
\begin{eqnarray}
   d \rho & = &  -\gamma D[\sigma_-]\rho dt + \sqrt{2 \eta \gamma} \left[ \sigma_- e^{i\theta} \rho + \rho \sigma_+  e^{-i\theta} - \langle \sigma_x\cos\theta + \sigma_y\sin\theta \rangle \rho \right] dW .
    \label{eqSME2}
\end{eqnarray}
We now rewrite this equation using the Bloch-sphere representation of the density matrix, ${\bf a} = (x,y,z)$, where $\rho = (1/2)(I + {\bf a}\cdot \boldsymbol{\sigma})$ and $\boldsymbol{\sigma} = (\sigma_x,\sigma_y,\sigma_z)$ and $I$ is the two-by-two identity matrix.  This gives 
\begin{eqnarray}
  d x  & = &  - \gamma x dt + \sqrt{2\eta\gamma} \left[ (1 + z) \cos\theta - x ( x \cos\theta + y\sin\theta ) \right] dW , \\ 
  d y  & = & - \gamma y dt + \sqrt{2\eta\gamma} \left[ (1 + z) \sin\theta - y ( x \cos\theta + y\sin\theta ) \right] dW , \\
  d z  & = & - 2 \gamma (1 + z) dt - \sqrt{2\eta\gamma} (1 + z) \left[ x \cos\theta + y \sin\theta \right] dW .
\end{eqnarray}
Defining the ``linear entropy'', $L$, by $L = 1 - \mbox{Tr}[\rho^2]$, and using the above equations we find that 
\begin{equation}
   d L = - \gamma \left\{ 2L [1 - \eta \left( x \cos\theta + y\sin\theta \right)] + (\eta-1)(1+z)^2  \right\} dt + \sqrt{8\eta\gamma} L \left( x \cos\theta + y\sin\theta \right) dW . 
\end{equation}
\end{widetext}
We wish to maximize the rate of decay of $L$ by adjusting the phase of the local oscillator, $\theta$, as the measurement proceeds. Inspection of the above equation makes it clear how to do this: we simply need to choose $\theta$ at each time so that $x \cos\theta + y\sin\theta = 0$. This not only maximizes the rate of decay of $L$, but also eliminates the stochastic terms in $dL$ and $dz$ so that the evolutions of both are   deterministic. This parallels the behavior of the rapid-purification algorithm in~\cite{rapidP}. When we choose $\theta$ at each time to maximize the rate of reduction of $L$, the evolution of $L$ becomes 
\begin{equation}
   \frac{dL}{dt} = - \gamma \left[ 2L + (\eta-1)(1 + z)^2  \right]  . 
\end{equation}
To achieve this we must continually adjust $\theta$ so that $\theta(t) =  \arg[y(t) - ix(t)]$.  With this choice of $\theta$ the equation for $z$ is simply $dz/dt = -2\gamma(1+z)$. We now take the initial state to be the maximally mixed single-qubit state $\rho(0) = I/2$. Solving for the evolution of $z$ in this case we have  
\begin{equation}
   z(t) = e^{-2\gamma t} - 1 ,
\end{equation}
and the equation of motion for the linear entropy becomes  
\begin{equation}
   \frac{dL}{dt} = - \gamma \left[ 2L  + (\eta - 1) e^{-4\gamma t} \right]  . 
\end{equation} 
Thus the evolution of the linear entropy, under the rapid-purification feedback 
algorithm is  
\begin{equation}
   L_{\mbox{\scriptsize fb}}(t) =  e^{-2\gamma t} \left[ \frac{1}{2}  
                                                   + \frac{1}{2} (1 - \eta )\left(1 - e^{-2\gamma t} \right) \right]  . 
   \label{fbL}
\end{equation} 

We now need to compare this with the evolution of the average value of the linear entropy in the absence of any feedback. (That is, when $\theta$ is fixed during the measurement.) Since we are 
treating the case when the initial state is maximally mixed, all choices for the fixed value of $\theta$ 
are equivalent, and so we will choose $\theta = 0$ for simplicity. When $\theta$ is fixed the evolution of $L$ is stochastic, and thus more complex. Nevertheless, for perfectly efficient detection ($\eta = 1$) the master equation Eq.(\ref{eqSME2}) is readily solved by using the linear form of the equivalent stochastic Schr\"{o}dinger equation (SSE), being~\cite{WisemanLinQ} (see also~\cite{GG,JK,JacobsSteck06}),
\begin{equation}
  d |\psi\rangle = \left[ -\gamma \sigma_+ \sigma_- dt + \sqrt{2\gamma} \sigma_- dW \right]  |\psi\rangle
  \label{linsse}
\end{equation}
The solution is 
\begin{equation}
   \rho(t) = \frac{V(t) \rho(0) V(t)^\dagger}{\mbox{Tr}[V(t)^\dagger V(t) \rho(0)]}  , 
\end{equation} 
where 
\begin{eqnarray} 
  V(t) & = &  e^{-\gamma \sigma_+ \sigma_- t} e^{R \sigma_- }  \nonumber \\
         & = & (e^{-\gamma t} \sigma_+ \sigma_- + \sigma_- \sigma_+) (1 + R \sigma_- ) 
\end{eqnarray} 
and $R$ is a random variable whose probability density at time $t$ is 
\begin{equation} 
   P(R,t) = \mbox{Tr}[V(t)^\dagger V(t) \rho(0)] \frac{e^{-R^2/ (2 \kappa)}}{\sqrt{2\pi \kappa}} , 
\end{equation} 
where we have defined $\kappa \equiv (1 - e^{-2\gamma t})$. When the initial state is the single-qubit maximally mixed state, $\rho(0) = I/2$, the solution is 
\begin{equation}
   \rho(t)  = \frac{\left[ e^{-2\gamma t} \sigma_+ \sigma_- +  e^{-\gamma t} \left( \sigma_+   + \sigma_- \right)  + \left(1 + R^2 \right) \sigma_- \sigma_+  \right] }{ 2 + R^2 - \kappa } 
\end{equation}
and 
\begin{equation}
  P(R,t) =  \frac{(2 + R^2 - \kappa)}{\sqrt{8\pi \kappa}} e^{-R^2/ (2 \kappa)} ,
\end{equation} 
The evolution of the average value of the linear entropy is then given by 
\begin{equation}
  \langle L(t) \rangle = \int_{-\infty}^\infty (1 - \mbox{Tr}[\rho(t)^2]) P(R,t) dR.
  \label{avL}
\end{equation} 
This integral cannot be solved analytically, and we will therefore evaluate it numerically. 

When $\eta$ is less than unity the SME is no longer equivalent to an SSE because it can increase the entropy of an initially pure state. Nevertheless, it turns out that it is possible to obtain an analytic solution to the SME by using the above technique of solving a linear SSE. As far as we know this method has not appeared in the literature to date, and so we describe it in detail in the appendix.  

\begin{figure}[t] 
\leavevmode\includegraphics[width=1.0\hsize]{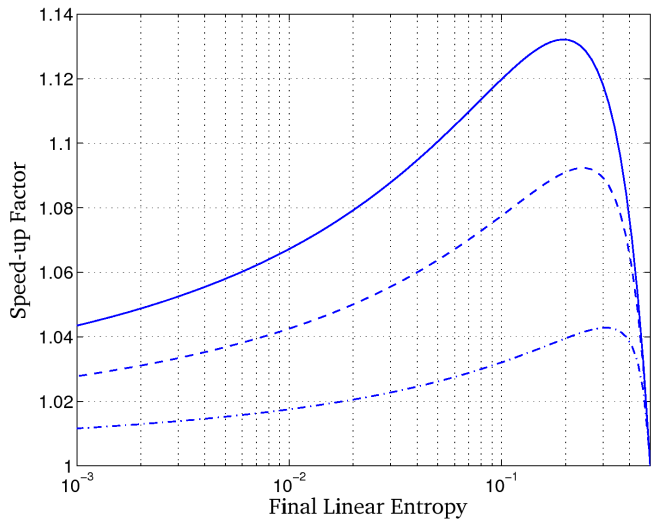} 
\caption{The speed-up factor in the time required to achieve a given final value of the average linear entropy, $\langle L \rangle$, afforded by the deterministic rapid-purification algorithm when the initial state of the optical qubit is completely mixed, as a function of $\langle L \rangle$. The various curves correspond to different values of the measurement efficiency $\eta$. Solid line: $\eta=1$; Dashed line: $\eta=0.8$; Dash-dot Line: $\eta=0.5$.} 
\label{fig1}
\end{figure}

When the initial state is $\rho(0) = I/2$ the solution for abitrary $\eta$ is 
\begin{eqnarray}
   \rho(t) & = & (2 \mathcal{N})^{-1} \left[ e^{-2\gamma t} \sigma_+ \sigma_- +  e^{-\gamma t} \left( \sigma_+   + \sigma_- \right) \right. \nonumber \\ 
              & & \left. \;\;\;\;\;\;\;\;\;\;\;\;\;\; + \left(1 + R^2 + [1-\eta]\kappa \right) \sigma_- \sigma_+  \right] , 
\end{eqnarray}  
where $\mathcal{N} = 1 + R^2/2 - \eta \kappa/2$ is the normalization constant. The probability density for the random variable $R$ is now 
\begin{eqnarray} 
   P(R,t) = \frac{2 + R^2 - \eta \kappa}{\sqrt{8\pi \eta \kappa }} e^{-R^2/(2\eta \kappa)} 
\end{eqnarray} 

We define the speed-up afforded by the rapid purification as the ratio of two times, $s=t_{\mbox{\scriptsize m}}/t_{\mbox{\scriptsize fb}}$. The first time, $t_{\mbox{\scriptsize m}}$, is that taken for $\langle L(t) \rangle$ to reach a given target value in the absence of feedback, and the second time, $t_{\mbox{\scriptsize fb}}$, is that taken for $L_{\mbox{\scriptsize fb}}(t)$ to reach the same target value. Using the expressions for the linear entropy in the two cases (Eqs. (\ref{fbL}) and (\ref{avL})) we plot this speed-up as a function of the target entropy in Figure~\ref{fig1}, and for three values of the detection efficiency $\eta$. We see that in the present case the speed-up factor reaches a peak and then decays back to unity as time increases. This is quite different behavior to that of the equivalent protocol for a measurement of an observable (a non-dissipative measurement), which tends to its maximum value as $t\rightarrow\infty$. 

\section{Analogues of the Wiseman-Ralph Rapid-Purification Protocol} 
\label{sec3} 

The Wiseman-Ralph protocol minimizes the average time, $\langle T \rangle$, taken to reach a given linear entropy, where the average is once again taken over all possible measurement records~\cite{Wiseman06x, Wiseman07}. In the case of a non-dissipative measurement on a single qubit, this is achieved by applying no feedback so long as there is no Hamiltonian evolution. We first ask, therefore, what is the speedup in $\langle T\rangle$ when we apply no feedback during optical homodyning over that provided by the deterministic feedback algorithm derived above? In this case we resort to  performing a numerical simulation of the measurement process (without feedback), and compare the resulting $\langle T \rangle$ (obtained by averaging over approximately five thousand trajectories) with that for the deterministic protocol as a function of the final linear entropy, L. For the latter one has a simple analytic expression for the time taken to reach the linear entropy $L$, given by solving Eq.(\ref{fbL}) for t, and this is 
\begin{equation}
   T = -\frac{1}{2\gamma}\ln\left[ \frac{1-\eta/2}{1-\eta} - \sqrt{\left(\frac{1-\eta/2}{1-\eta}\right)^2 - \frac{2L}{1-\eta}}\right] 
   \label{Tdet}
\end{equation}
were for $\eta=1$ this reduces to $T = -\ln(2L)/(2\gamma)$. 

\begin{figure}[t] 
\leavevmode\includegraphics[width=1.0\hsize]{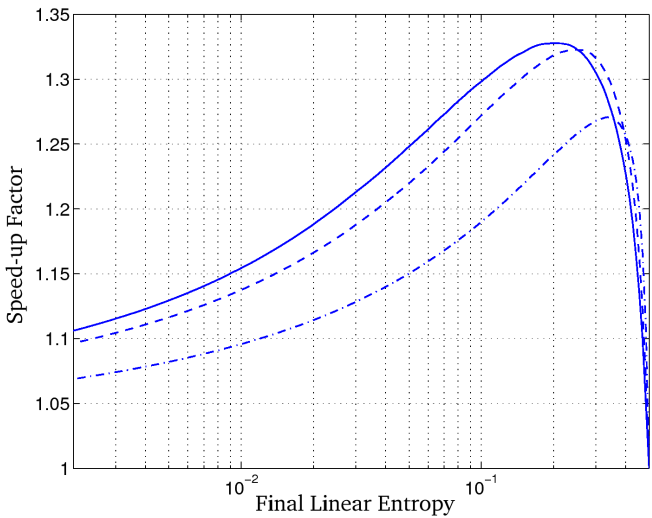} 
\caption{The speed-up in the average time taken to reach a given linear entropy $L$ for no feedback over that of the deterministic feedback algorithm presented in Section~\ref{sec2}. The various curves correspond to different values of the measurement efficiency $\eta$. Solid line: $\eta=1$; Dashed line: $\eta=0.8$; Dash-dot Line: $\eta=0.5$.} 
\label{fig2}
\end{figure}

We plot the resulting speed-up factor in Figure~\ref{fig2} for three values of the measurement efficiency $\eta$. As expected there is a speedup, although once again this speed-up vanishes in the long-time limit. 

The absence of feedback is not, however, the equivalent of the Wiseman-Ralph protocol, because in our case it does not necessarily provide the minimum value of $\langle T \rangle$. Without loss of generality let us choose the phase of the local oscillator to be $\theta=0$, which is the homodyne equivalent of measuring a qubit in the $\sigma_x$ basis. Examining the equations of motion for $z$ and $x$ in this case we see that in the absence of feedback the Bloch vector does not remain along the $x$-axis, as would be true of the W-R protocol for a $\sigma_x$ measurement, but in the $x-z$ plane. This is, however, the closest we can come to the W-R protocol if we apply feedback only to the phase of the local oscillator, $\theta$; it keeps the $x-y$ component of the Bloch vector  aligned with the measured quadrature. Modifying the local oscillator is much simpler to implement experimentally that applying unitary operations to the optical mode, and for this reason is of most interest to us here. 

If we could implement feedback consisting of unitary operations on the state of the optical mode, then we could rotate the qubit during the measurement so as to keep the Bloch vector aligned with the basis of the measurement. For $\theta=0$ the measurement basis is the $x$-basis, so the result is $z=0$, $y=0$ and the evolution is given by a single differential equation for $x$, being 
\begin{equation}
   dx = -(1-\eta)\gamma x dt + \sqrt{2\eta\gamma}(1-x^2) dW .
\end{equation}
When $\eta=1$ the deterministic term vanishes, and the equation is essentially identical to that describing the Wiseman-Ralph protocol. In this case one can obtain an analytic solution for the average time to reach a given linear entropy $L$~\cite{Wiseman06x,gardiner}. This gives 
\begin{equation}
  \langle T \rangle =  \frac{\sqrt{1-2L}}{4\gamma}\ln\left[\frac{1 + \sqrt{1-2L}}{1 - \sqrt{1-2L}} \right] 
\end{equation}
In the limit $L\rightarrow 0$ (equivalently, $t\rightarrow \infty$) this reduces to $\langle T \rangle \approx -\ln(L/2)/(4\gamma)$. Using this expression for $\langle T \rangle$, and that given in Eq.(\ref{Tdet}), we can immediately calculate the speedup factor for this protocol over the protocol in Section~\ref{sec2}. This is 
\begin{equation} 
  s_\infty \equiv  \lim_{t\rightarrow\infty}(T/\langle T \rangle)= 1/2.  
\end{equation}
Thus, unlike the previous two protocols, the speed-up factor for this protocol does not decay to unity as $t\rightarrow\infty$. 

\begin{figure}[t] 
\leavevmode\includegraphics[width=1.0\hsize]{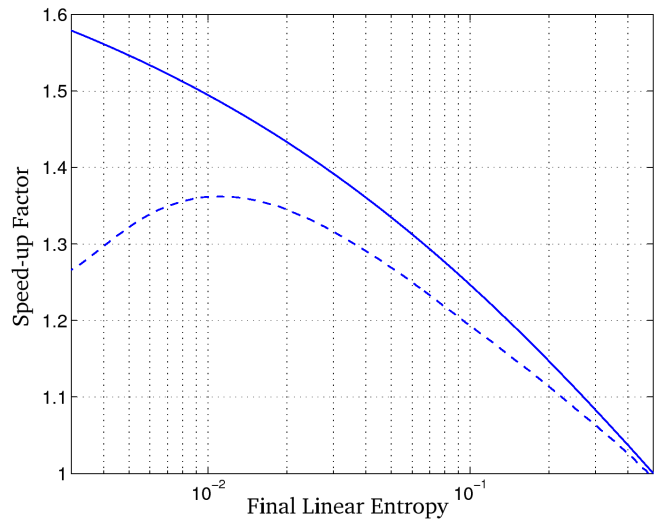} 
\caption{The speed-up in the average time taken to reach a given linear entropy $L$ for an analogue of the Wiseman-Ralph feedback protocol over that of the deterministic feedback algorithm presented in Section~\ref{sec2}. The two curves correspond to different values of the measurement efficiency $\eta$. Solid line: $\eta=1$; Dashed line: $\eta=0.95$.} 
\label{fig3}
\end{figure}

Since we cannot obtain an analytic solution for $\eta<1$, we calculate the speed-up factor for $\eta=0.8$ by obtaining $\langle T \rangle$ using a numerical simulation and averaging over approximately five thousand trajectories. We plot the speed-up factor as a function of the final linear entropy for three values of $\eta$ in Figure~\ref{fig3}. From this we see that the speed-up for this algorithm is quite a lot more sensitive to the measurement efficiency than the previous two. 

\section{Conclusion}
\label{conc}
We have considered applying feedback to the homodyne detection of a single optical qubit so as to change the rate at which the system is purified (so called ``rapid-purification feedback algorithms''). We have shown that there exists a feedback algorithm that increases the rate at which the {\em average purity} increases, and like its non-dissipative counterpart this results in a deterministic evolution for the purity of the system. Unlike its non-disspative analogue, the speed-up provided by this protocol reaches its maximum value at a finite time, decaying to unity as $t\rightarrow\infty$. We also found that the speed-up remains for measurement efficiencies well below unity, although the speed-up decreases as the efficiency drops. 

We have also examined the behavior of the {\em average time} taken to reach a fixed purity for different feedback algorithms. We found, as is true for non-dissipative measurements, that the protocol presented in Section~\ref{sec2} is slower than without feedback, as is the case for non-dissipative measurements. We also pointed out that the measurement without feedback is not the closest analogue of the Wiseman-Ralph feedback protocol for homodyne detection. We considered a more closely analogous algorithm in which the Bloch vector is aligned with the effective direction of the measurement, although this feedback cannot be implemented by merely changing the phase of the local oscillator. We found that this protocol behaves much more like those for a non-dissipative measurement, in that the speed-up factor increases monotonically and tends to a value of two in the long-time limit. This protocol is, however, more sensitive to noise than the previous protocols. The above results show that it should be feasible to demonstrate rapid-purification protocols in an optical setting. 

\section*{Acknowledgments} 
The authors would like to acknowledge support from the Army Research Office and the Disruptive Technologies Office. The authors also acknowledge the use of the supercomputer facilities in the College of Science and Mathematics at UMass Boston. 

\appendix

\section{Solving the SME for inefficient detection}  

We first note that a master equation that describes inefficient detection is equivalent to a master equation containing two simultaneous measurements, where the observer has access to only one, and must average over the results of the other~\cite{JacobsSteck06}. Our method is then to solve the Stochastic Schr\"{o}dinger equation equivalent to the SME with two measurements (by using the method of linear quantum trajectories~\cite{WisemanLinQ, GG, JK, JacobsSteck06}), and then take the average over the second measurement at the end to obtain the solution for inefficient detection. It turns out that the resulting integrals are straightforward and give a fully analytic solution. The SME Eq.(\ref{SMEhom}) is thus equivalent to the linear SSE~\cite{JacobsSteck06}) 
\begin{equation}
   d |\psi\rangle = \left[ -\gamma a^\dagger a dt + \sqrt{2\eta\gamma} a dW + \sqrt{2(1-\eta)\gamma} a dV \right]  |\psi\rangle
\end{equation} 
where $dW$ and $dV$ are independent Gaussian noise sources so that $dWdV = 0$. The observer has access to the measurement record corresponding the measurement associated with $dW$, and thus must ultimately average over $dV$. We obtain the evolution operator which solves this equation by using the method given in reference~\cite{JK}, and this is  
\begin{equation}
   V(t, R, Q)  = e^{-\gamma a^\dagger a t} e^{ \kappa a^2}e^{ a R} e^{a Q}
\end{equation}
where 
\begin{eqnarray}
   R & = &  \sqrt{2\eta \gamma} \int_0^t e^{-2\gamma s}  dW(s)   \label{eqR} \\
   Q & = &   \sqrt{2(1 - \eta)\gamma}\int_0^t e^{-2\gamma s}  dV(s)   \label{eqQ}
\end{eqnarray}
and $\kappa\equiv (1 - e^{-2\gamma t})$. The probability densities for $R$ and $Q$ resulting from the above stochastic integrals are Gaussian, with mean zero and variances $V_R = \eta \kappa$ and $V_Q = (1-\eta)\kappa$. We will denote these Gaussian densities by $G(R)$ and $H(Q)$, respectively. 

For an initial state $\rho(0)$, the solution is thus 
\begin{equation}
   \rho(t, R, Q)  =  \frac{V \rho(0) V^\dagger }{ \mathcal{N}}
\end{equation}
where $\mathcal{N} = \mbox{Tr}[V^\dagger V \rho(0)]$ is the normalization. The true joint probability density for $R$ and $Q$ is given by the product of the Gaussian densities $G(R)$ and $H(Q)$, multiplied by $\mathcal{N}$. That is  
\begin{equation}
   P(R,Q,t) =  \mbox{Tr}[V^\dagger V \rho(0)] G(R) H(Q) .
\end{equation}
To obtain the solution to the inefficient SME we must average over the $Q$ keeping $R$ fixed. This solution is therefore 
\begin{eqnarray}
   \sigma(R,t) & = &  \int_{-\infty}^\infty  \rho (R,Q,t) P(Q|R)  dQ  \nonumber \\
                     & = &  \frac{1}{P(R)} \int_{-\infty}^\infty   \rho (R,Q,t) P(R,Q)  dQ  \nonumber \\
                     & = & \frac{G(R)}{P(R)} \int_{-\infty}^\infty  V \rho(0) V^\dagger  H(Q)  dQ  \nonumber \\ 
                     & = & \frac{1}{\mathcal{M}} \int_{-\infty}^\infty  V \rho(0) V^\dagger  H(Q)  dQ 
\end{eqnarray}
where $\mathcal{M}$ is merely the normalization. From this we see that we need only perform an integration over the Gaussian density for $Q$, which is straightforward. 


\end{document}